# CONDITION-BASED MAINTENANCE USING SENSOR ARRAYS AND TELEMATICS

Gopalakrishna Palem

Symphony-Teleca Corporation, Bangalore, India
Gopalakrishna.Palem@Yahoo.com

## ABSTRACT

*Emergence of uniquely addressable embeddable devices has raised the bar on Telematics capabilities. Though the technology itself is not new, its application has been quite limited until now. Sensor based telematics technologies generate volumes of data that are orders of magnitude larger than what operators have dealt with previously. Real-time big data computation capabilities have opened the flood gates for creating new predictive analytics capabilities into an otherwise simple data log systems, enabling real-time control and monitoring to take preventive action in case of any anomalies. Condition-based-maintenance, usage-based-insurance, smart metering and demand-based load generation etc. are some of the predictive analytics use cases for Telematics. This paper presents the approach of condition-based maintenance using real-time sensor monitoring, Telematics and predictive data analytics.*

## 1. INTRODUCTION

Maintenance, considered often as a non-value add function, is always under a constant pressure from the top management to contribute more for costs reduction, keep the machines in excellent working condition, all the while satisfying the stringent safety and operational requirements. Towards this end manufacturers and operators usually employ various maintenance strategies, all of which and can be broadly categorized as below:

- Corrective Maintenance
- Preventive Maintenance
- Predictive Maintenance

Corrective maintenance is the classic *Run-to-Failure* reactive maintenance that has no special maintenance plan in place. The machine is *assumed* to be fit unless proven otherwise.

- Cons:
  - High risk of collateral damage and secondary failure
  - High production downtime
  - Overtime labour and high cost of spare parts

- Pros:
  - Machines are not over-maintained
  - No overhead of condition monitoring or planning costs

Preventive maintenance (PM) is the popular *periodic maintenance* strategy that is actively employed by all manufacturers and operators in the industry today. An optimal breakdown window is pre-calculated (at the time of component design or installation, based on a wide range of models describing the degradation process of equipment, cost structure and admissible maintenance etc.), and a preventive maintenance schedule is laid out. Maintenance is carried-out on those periodic intervals, *assuming* that the machine is going to break otherwise.

- Cons:

    - Calendar-based maintenance: Machines are repaired when there are no faults
    - There will still be unscheduled breakdowns

- Pros:

    - Fewer catastrophic failures and lesser collateral damage
    - Greater control over spare-parts and inventory
    - Maintenance is performed in controlled manner, with a rough estimate of costs well-known ahead of time

Predictive Maintenance, also known as *Condition-based maintenance* (CBM) is an emerging alternative to the above two that employs predictive analytics over real-time data collected (streamed) from parts of the machine to a centralized processor that detects variations in the functional parameters and detects anomalies that can potentially lead to breakdowns. The real-time nature of the analytics helps identify the functional breakdowns long before they happen but soon after their potential cause arises.

- Pros:
    - Unexpected breakdown is reduced or even completely eliminated
    - Parts are ordered when needed and maintenance performed when convenient
    - Equipment life is maximized
- Cons:
    - High investment costs
    - Additional skills might be required

Condition-based maintenance differs from schedule-based maintenance by basing maintenance need on the actual condition of the machine rather than on some pre-set schedule.

For example, a typical schedule-based maintenance strategy demands automobile operators to change the engine oil, say after every 3,000 to 5,000 Miles travelled. No concern is given to the actual condition of vehicle or performance capability of the oil.

If on the other hand, the operator has some way of knowing or somehow measuring the actual condition of the vehicle and the oil lubrication properties, he/she gains the potential to extend the vehicle usage and postpone oil change until the vehicle has travelled 10,000 Miles, or perhaps pre-pone the oil change in case of any abnormality.

In the following sections we look into what constitutes a condition-based maintenance solution, the steps involved in implementing one and an overall solution methodology.

## 2. CONDITION-BASED MAINTENANCE

Predictive analytics in combination with sensor based telematics provides deep insights into the machine operations and full functionality status – giving rise to optimal maintenance schedules with improved machine availability.

Underlying schedule-based maintenance is the popular belief that machine failures are directly related to machine operating age, which studies indicate not to be true always. Failures are not always linear in nature. Studies indicate that 89% of the problems are random with no direct relation to the age [2]. Table 1 showcases some of these well-known failure patterns and their conditional probability (Y-axis) with respect to Time (X-axis).

Table 1. Failure Conditional Probability Curves

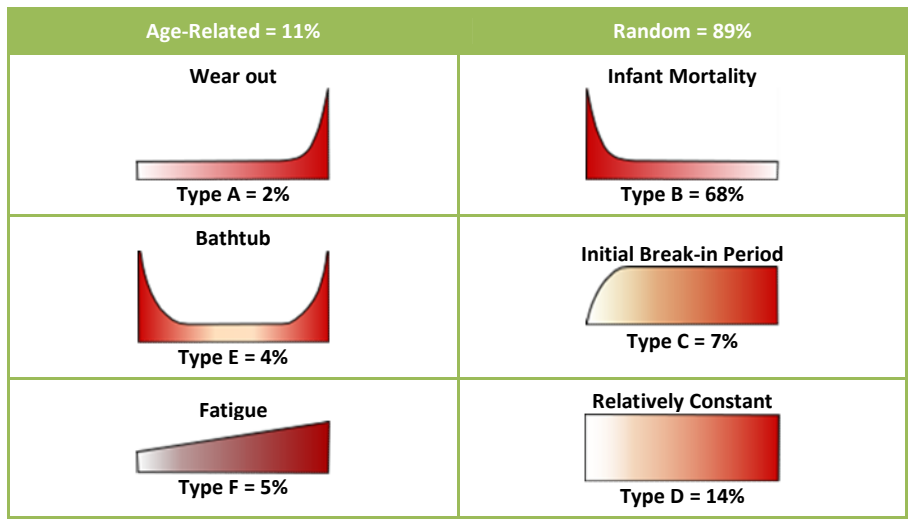

Complex items frequently demonstrate some infant mortality, after which their failure probability either increases gradually or remains constant, and a marked wear-out age is not common. Considering this fact, the chance of a schedule-based maintenance avoiding a potential failure is low, as there is every possibility that the system can fail right after a scheduled maintenance. Thus, preventive maintenance imposes additional costs of repair. Condition-based maintenance reduces such additional costs by scheduling maintenance if and only when a potential breakdown symptom is identified.

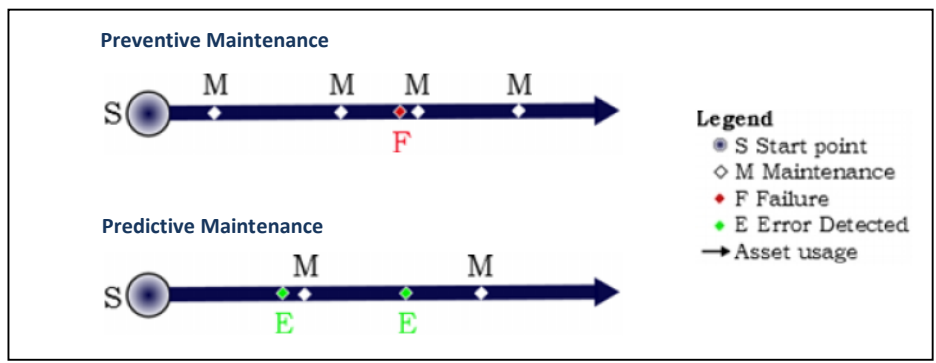

Figure 1. Condition-based maintenance schedules are flexible and cost-saving

However, the costs of monitoring equipment and monitoring operations should not exceed the original asset replacement costs; lest the whole point of condition-based maintenance becomes moot. Internal studies conducted with our customers have estimated that a properly functioning CBM program can provide savings of 8% to 12% over the traditional maintenance schemes. They indicated the following industrial average savings resultant from initiation of a functional predictive maintenance program:

- Reduction in maintenance costs: 25% to 30%
- Elimination of breakdowns: 70% to 75%
- Reduction in equipment or process downtime: 35% to 45%
- Increase in production: 20% to 25%

Apart from the above, improved worker and environment safety, increased component availability, better product quality etc. are making more and more manufacturers and operators embrace CBM based management solutions.

## 2.1. SOLUTION ENABLERS

A Condition-based maintenance management (CBMM) solution is enabled by three major technology enhancements over a traditional maintenance solution:

1. Remote Sensor Monitoring & Data Capturing
2. Real-time Stream Processing of Sensor Data
3. Predictive Analytics

CBMM solutions essentially operate by having sensors attached to remote assets (mobile or stationary) that send continuous streams of data about the assets' operational conditions to a monitoring station that then analyses them in real-time using predictive analytic models and detects any problems in the behaviour or state of the asset. Once a problem is detected, appropriate pre-configured action is taken to notify the operator or manufacture for corrective action. The monitoring station in question can be on the same network as that of the sensors or it could be in a remote location far away from them, connected through wide area networks or satellite networks.

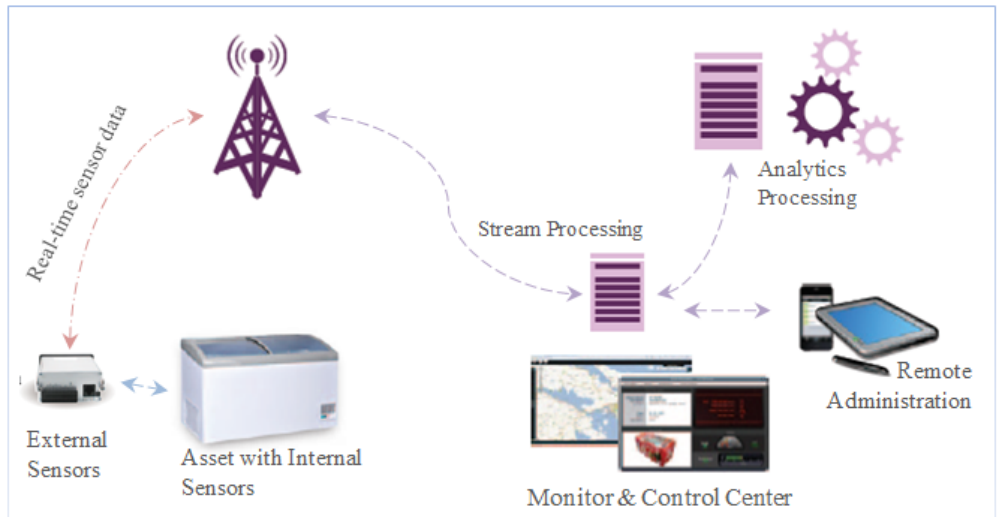

Figure 2. Condition-based maintenance using sensor arrays and telematics

Nature of the sensors being monitored, frequency of the data getting collected and precision of the analytic models being used – all affect the quality of the prediction results. Thus, it is imperative that manufacturers and operators define all these parameters with utmost care while deploying a CBMM system. This, however, entails a thorough understanding of the system under operation and expects a clear-cut answer as to what is being monitored and what is expected out of such monitoring. Some of the questions that can help manufacturers and operators along those lines are:

For monitoring:
- Which parts of the system or asset are expected to be monitored?
- What type of data is expected to be collected and which type of sensors give such data? For example, visual data, thermal data etc.
- What is the expected frequency for the data collection?
- How should any failures in the sensors be handled?

For real-time stream collection and processing at the monitoring station:
- What is the acceptable data processing latency?
- How to deal with imperfections in the received data? For example, a faulty sensor sending incorrect data
- What should be done with the collected data after processing?

For Analytics sub-system:
- Which analysis technique accurately models the asset/system behaviour?
- What is the definition of acceptable behaviour and anomaly?
- What should be the response in case of any anomaly detection?
- What should be the reasonable timeframe between anomaly detection and corrective action?
- How to deal with situations where there are multiple anomalies detected at the same time?

Generic and security related questions:
- Who should be allowed to access the collected data and analysis results?
- What is the change management process required in case one wants to tune the tracking and analysis parameters?

The following section briefly summarizes some of the standard methods we use in the CBMM systems we deploy for our customers and can help in answering above questions.

## 2.2. THE METHODOLOGY

The primary component in a condition-based maintenance management solution is a *sensor array* and the measurements it provide. Some of the widely used measurement techniques in the industry are:

- Temperature Measurement: Thermal indicators, such as temperature-sensitive paint, thermography etc., help detect potential failures arising out of temperature changes in the equipment. Excessive mechanical friction, degraded heat transfer, poor electrical connections are some of the problems that can be detected with this type of measurement.

| Method | Description | Applications |
| --- | --- | --- |
| Point Temperature | A thermocouple or RTD | Can be used on all accessible surfaces |
| Area Pyrometer | IR radiation measured from a surface, often with laser sight | Good for walk around temperature checks on machines and panels |

| Temperature Paint | Chemical indicators calibrated to change colors at specified temperature | Works great for inspection rounds |
| Thermography | Handheld still or video camera sensitive to emitted IR | Best for remote monitoring. Requires good training |

- Dynamic Monitoring:  Spectrum analysis, shock pulse analysis are some of  the dynamic monitoring methods that measure and analyse energy emitted from mechanical equipment in the form of waves, vibration, pulses and acoustic effects. Wear and tear, imbalance, misalignment and internal surface damage are some of the problems that can be detected with this type of measurement.

| Method | Description | Applications |
| --- | --- | --- |
| ISO Filtered velocity | 2Hz-1kHz filtered velocity | A general condition indicator |
| SPM | Carpet and Peak related to demodulation of sensor resonance around 30kHz | Single value bearing indicator method |
| Acoustic Emission | Distress & dB, demodulates a 100kHz carrier sensitive to stress waves | Better indicator than ISO velocity, without the ISO comfort zone |
| Vibration Meters | Combine velocity, bearing and acceleration techniques | ISO Velocity, envelope and high frequency acceleration give best performance |
| 4-20mA sensors | Filter data converted to DCS/PLC compatible signal | Useful to home-in on specific problems by special order |

- Fluid Analysis: Ferrography, particle counter testing are some of the fluid analysis methods performed on different types of oils, such as lubrication, hydraulic, insulation oil etc., to identify any potential problems of wear and tear in the machines. Machine degradation, oil contamination, improper oil consistency, oil deterioration are some of the problems that can be detected with this method. The main areas of analysis in this are:

  - Fluid physical properties: Viscosity, appearance
  - Fluid chemical properties: TBN, TAN, additives, contamination, % water
  - Fluid contamination: ISO cleanliness, Ferrography, Spectroscopy, dissolved gases
  - Machine health: Wear metals associated with plant components
- Corrosion Monitoring: Methods such as Coupon testing, corrometer testing help identify the extent of corrosion, corrosion rate and state (active/passive corrosion) for the materials used in the asset.
- Non-destructive Testing: Involves using non-destructive methods, such as X-Rays, ultrasonic etc., to detect any potential anomalies arising internal to the asset structure. Most of these tests can be performed while the asset is online and being used.
- Electrical testing and Monitoring: High potential testing, power signal analysis are some of the prominent electrical condition monitoring mechanisms that try to identify any changes in the system properties, such as resistance, conductivity, dielectric strength and potential. Electrical insulation deterioration, broken motor rotor bars and shorted motor stator lamination etc. are some of the problems that can be detected with this type of mechanism.

- Observation and Surveillance: Visual, audio and touch inspection criteria are some of the surveillance condition monitoring techniques based on the human sensory capabilities. They act as supplement to other condition-monitoring techniques and help detect problems such as loose/worn parts, leaking equipment, poor electrical and pipe connections, stream leaks, pressure relief valve leaks and surface roughness changes etc.

Once the appropriate measurement mechanisms are in place, the next step is the event definition phase: to define what constitutes acceptable system behaviour and what is to be considered as anomaly. It is useless to put costly monitoring equipment in place, without knowing what to expect out of it. Expert opinion and judgment (such as manufacturer's recommendations), published information (such as case studies), historical data etc. are some of the good sources that can help in this task. The definition of anomaly should be unambiguous and easy to detect. If the cost of anomaly detection far exceeds the costs of consequences of that anomaly, then it is not a valid scenario for implementing CBMM system.

The next step that follows the event definition phase is, determining event inspection frequency. Frequency of any of form of condition-based-maintenance is based on the fact that most failures do not occur instantaneously, and that it is often possible to detect them during their final stages of deterioration. If evidence can be found that something is in the final stages of failure, it is possible to take action for preventing it from failing completely and/or avoid the consequences.

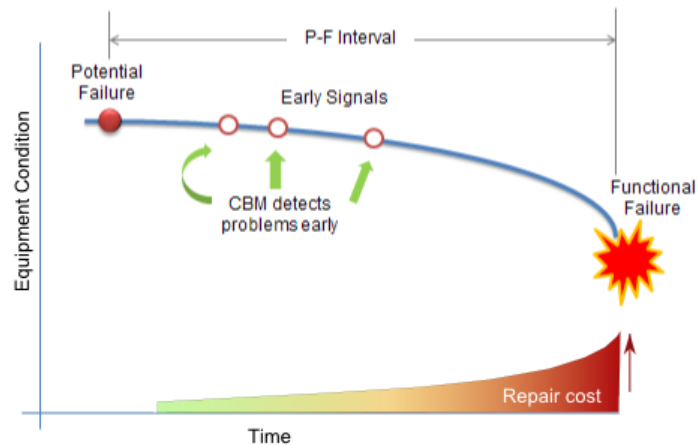

Figure 1. Measurement Frequencies can be determined with P-F Interval

The failure behaviour typically exhibited by majority of the systems in operation is as showcased in Figure 3. During operation, over a period of time, the systems enter a phase of potential failure (P), and start displaying few early signs of wear & tear and other stressful behaviours that if neglected finally lead to full functional failure (F). For most of the systems the time interval between the potential failure point (P) and full function failure (F) is large enough to allow detection and prevention of the failure.

This time gap between P and F is what is popularly known as the *P-F Interval* in the literature, and any cost-effective maintenance strategy should try to maximize on it. The P-F Interval could be in hours or days, or even weeks or months, based on the complexity of the system and the unit of measurement - for it is not uncommon to see the P-F Interval being measured in non-time units, such as stop-start cycles or units of output etc. Based on the failure mode and the unit of measurement, the P-F Interval can end up varying from fractions of a second to several decades (on temporal scale).

Be whatever the unit of measurement and the P-F interval, a successful CBMM system should be capable of detecting the early signals after P and respond to them long before F. The response action typically consists of multiple steps (as laid out below) and should all be accompanied within the P-F interval.

1. Analysing the root-cause based on detected early signals
2. Planning corrective action based on the analysed root-cause
3. Organizing the resources to implement the laid out plan
4. Actual implementation of the corrective action plan

The amount of time needed for these response actions usually vary, from a matter of hours (e.g. until the end of operating cycle or end of shift), minutes (e.g. to clear people from a failing building), to weeks or even months (e.g. until a major shutdown).

Thus, it is a common practice to use the inspection interval to be half the P-F interval. This will ensure that there is at-least half the P-F interval remaining after the potential-failure detection for corrective action plan. However, it should be noted that most of the times earlier the corrective action plans are implemented, lower the cost – in which cases, some other smaller fraction of P-F interval can be used as the inspection interval, so that potential problems can be detected as early as possible and rectified.

However, an important point to be remembered is P-F interval is not an easy metric to be computed. It varies from asset type to asset type, environment to environment and even from one asset to another with in the same asset type (based on its previous fault history and working conditions). Understanding the failure patterns, and identifying the class of pattern to which the asset belongs, its past fault history, manufacturer's recommendations, operating conditions, expert judgment etc. are some of the sources that can help in arriving at an accurate P-F interval for any given asset/system.

At each inspection interval, the CBMM system collects data from sensors and uses one of the following methods to determine the condition of the asset being monitored:

- Trend Analysis: Reviews the data to find if the asset being monitored is on an obvious and immediate downward slide toward failure. Typically a minimum of three monitoring points are recommended for arriving at a trend accurately as a reliable measure to find if the condition is deprecating linearly.
- Pattern recognition: Decodes the causal relations between certain type of events and machine failures. For example, after being used for a certain product run, one of the components used in the asset fails due to stresses that are unique to that run
- Critical range and limits: Tests to verify if the data is within a critical range limit (set by professional intuition)
- Statistical process analysis: Existing failure record data (retrieved from warranty claims, data archives and case-study histories) is driven through analytical procedures to find an accurate model for the failure curves and the new data is compared against those models to identify any potential failures.

Based on the failure mode and asset class the right method for the prediction can vary. For example, assets that fall into type E class (bathtub pattern) usually benefit from Weibull distribution, while split system approach is used for complex systems with multiple sub-systems.

Stream processing the arriving data can help build the trend analysis and critical range limits, but to accurately process pattern recognition and statistical model building methods, past history data is as important as the new arriving data. Thus, typically CBM management systems should keep record of old data for some reasonable amount of time before they are archived or destroyed. This time period varies from domain to domain and may even be regulated by local country laws. For example, financial fraud records may need to be kept active for longer time, in the range of 7 to 15 years per se, while flight records generated from airplane internal sensors are typically discarded after the journey completion (primarily due to it being voluminous, though this trend could soon change as the big data warehousing gets more prominent).

Another reason the old data streams become important is to identify any potential outliers in the streamed-in data from the sensors. While monitoring for the faults in the assets, it is possible that the sensor that is taking the readings, being a machine itself, could fail and start sending faulty records. Intelligent CBM management systems capable of detecting such outliers will try to isolate these faulty sensors and notify the appropriate personal for corrective action, or substitute it with proper estimated data based on previous records. In either case, human inspection is as much necessary as a completely automated monitoring system – for automation only complements the human surveillance efforts, not replace them. Thus, many automated monitoring systems provide a way for manual override for configurable parts of their functionality.

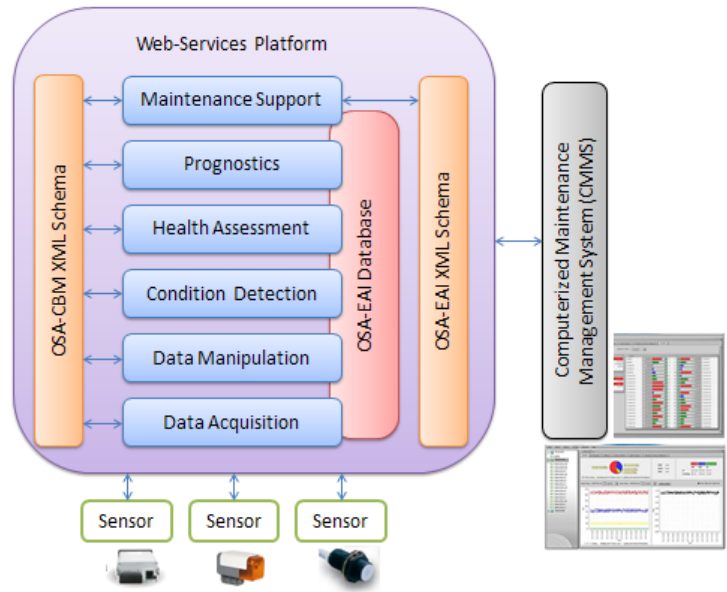

Figure 2. Condition-based maintenance management system architecture

Typically a CBM management solution will have an administrative console that lets the operators define and update various parameters, such as critical limits, response notifications, default corrective actions etc. Advanced management systems also allow the administrators to access the monitoring solution functionality remotely through web UI and mobile UI, capable of sending periodic digests weekly or monthly for pre-configured stake-holders reporting the status of the asset being monitored on a regular basis. With architectures like the one shown in Figure 4 and common interface standards such as IEEE 1451, IEEE 1232, MIMOSA and OSA-CBM, advanced management systems integrations become possible among disparate software and hardware components from different vendors, all working hard together just for one single purpose – to provide the operators maximum usage out of their assets.

## 3. CONCLUSIONS

Condition-based maintenance is based on the principle of using real-time data to prioritize and optimize maintenance resources. Such a system will determine the equipment's health, and act only when maintenance is actually necessary. Telematics development (such as IPV6, 3G and 4G LTE) in the recent times in combination with big-data real-time stream analytics is opening new opportunities for manufacturers and asset owners to save costs and optimize resource usage in innovative ways. Condition-based maintenance management systems built around real-time sensor monitoring and telematics technologies offer flexibility and cost-savings in terms of providing greater control over when to perform the maintenance, which parts to pre-order and how the optimally schedule the labour.

## REFERENCES

[1]     Weibull, W, (1951) "A statistical distribution function of wide applicability", Journal of Applied Mechanics-Trans. ASME, Vol. 18, No. 3: pp. 293–297.

[2]     Nowlan, F. Stanley, and Howard F. Heap, (1978) *Reliability-Centred Maintenance. Department of Defense*, Report Number AD-A066579 Washington, D.C. 1978.

[3]     Pérez, Angel Torres; Hadfield, Mark. (2011). "Low-Cost Oil Quality Sensor Based on Changes in Complex Permittivity." Sensors 11, no. 11: 10675-10690.

[4]     J. CUENA, M. MOLINA, (2000) "The role of knowledge modelling techniques in software development: a general approach based on a knowledge management tool", International Journal of Human-Computer Studies, Vol. 52, No. 3, pp. 385-421

[5]     G. Abdul-Nour, H. Beaudoin, P. Ouellet, R. Rochette, S. Lambert, (2000) "A reliability based maintenance policy; a case study", Computers & Industrial Engineering, Vol. 35, No. 3-4, pp. 591-594

[6]     Hansen, T., Dirckinck-Holmfeld, L., Lewis, R., & Rugelj, J. (1999). Using telematics to support collaborative knowledge construction. *Collaborative learning: Cognitive and computational approaches*, 169-196.

**Authors**

Gopalakrishna Palem is a Corporate Technology Strategist specialized in Distributed Computing technologies and Cloud operations. During his 12+ year tenure at Microsoft and Oracle, he helped many customers build their high volume transactional systems, distributed render pipelines, advanced visualization & modelling tools, real-time dataflow dependency-graph architectures, and Single-sign-on implementations for M2M telematics. When he is not busy working, he is actively engaged in driving open-source efforts and guiding researchers on Algorithmic Information Theory, Systems Control and Automata, Poincare recurrences for finite-state machines, Knowledge modelling in data-dependent systems and Natural Language Processing (NLP).